\documentclass[conference]{IEEEtran}
\usepackage{url}
\usepackage[utf8]{inputenc}
\usepackage[table,xcdraw]{xcolor}
\usepackage{amsmath}
\usepackage{amsthm}
\usepackage{amssymb}
\usepackage{xfrac}
\usepackage{cite}
\usepackage{comment}
\usepackage[acronyms,nonumberlist,nopostdot,nomain,nogroupskip]{glossaries}
\usepackage{tablefootnote}
\usepackage{booktabs}
\usepackage{balance}

\usepackage{bm}
\usepackage{tikz}
\usepackage{pgfplots}
\pgfplotsset{compat=newest}
\pgfplotsset{plot coordinates/math parser=false}
\newlength\fheight
\newlength\fwidth
\usetikzlibrary{plotmarks,decorations,backgrounds,calc,spy}
\usetikzlibrary{decorations.markings}
\usepgfplotslibrary{patchplots,groupplots}
\usepackage{tikzscale}
\usepackage{hyperref}
\usepackage{multirow}
\usepackage[inline]{enumitem}
\usepackage{pgfplotstable} 
\usepackage[font=footnotesize]{subcaption}
\usepackage[font=footnotesize]{caption}

\usepackage{mathtools}

\definecolor{desireRed}{RGB}{230,57,60}%
\definecolor{darkPurple}{RGB}{59,31,43}%
\definecolor{springGreen}{RGB}{37,223,145}%
\definecolor{queenBlue}{RGB}{69,123,157}%
\definecolor{spaceCadet}{RGB}{29,53,87}%

\definecolor{primaryColor}{HTML}{F06449}
\definecolor{secondaryColor}{HTML}{5BC3EB}
\definecolor{tertiaryColor}{HTML}{36382E}

\newcommand{\oldversion}[1]{{\color{red}WAS: #1}}
\renewcommand{\oldversion}[1]{}

\newtheorem{thm}{Theorem}

\usepackage{bbm}
\IEEEoverridecommandlockouts % To allow \thanks{} command to work

\begin{document}

\newacronym{ca}{CA}{Carrier Aggregation}
\newacronym{3gpp}{3GPP}{3rd Generation Partnership Project}
\newacronym{5g}{5G}{5th generation}
\newacronym{5gc}{5GC}{5G Core}
\newacronym{adc}{ADC}{Analog to Digital Converter}
\newacronym{afbw}{AFBW}{Average Fading Bandwidth}
\newacronym{aimd}{AIMD}{Additive Increase Multiplicative Decrease}
\newacronym{am}{AM}{Acknowledged Mode}
\newacronym{amc}{AMC}{Adaptive Modulation and Coding}
\newacronym{aoa}{AoA}{Angle of Arrival}
\newacronym{aoi}{AoI}{Age of Information}
\newacronym{aod}{AoD}{Angle of Departure}
\newacronym{ap}{AP}{Access Point}
\newacronym{arvr}{AR/VR}{augmented/virtual reality}
\newacronym{af}{AF}{Amplify-and-Forward}
\newacronym{hpbw}{HPBW}{Half-Power Beamwidth}
\newacronym{rr}{RR}{regenerative relay}
\newacronym{aqm}{AQM}{Active Queue Management}
\newacronym{awgn}{AGWN}{Additive White Gaussian Noise}
\newacronym{balia}{BALIA}{Balanced Link Adaptation}
\newacronym{bdp}{BDP}{Bandwidth-Delay Product}
\newacronym{ber}{BER}{Bit Error Rate}
\newacronym{bf}{BF}{Beamforming}
\newacronym{bwp}{BWP}{Bandwidth Part}
\newacronym{cad}{CAD}{Computer-Aided Design}
\newacronym{cc}{CC}{Congestion Control}
\newacronym{cdf}{CDF}{Cumulative Distribution Function}
\newacronym{cir}{CIR}{Channel Impulse Response}
\newacronym{cn}{CN}{Core Network}
\newacronym{csma}{CSMA}{carrier sensing multiple access}
\newacronym{cp}{CP}{Control Plane}
\newacronym{cqi}{CQI}{Channel Quality Information}
\newacronym{crs}{CRS}{Cell Reference Signal}
\newacronym{csirs}{CSI-RS}{Channel State Information - Reference Signal}
\newacronym{dc}{DC}{Dual Connectivity}
\newacronym{dce}{DCE}{Direct Code Execution}
\newacronym{dci}{DCI}{Downlink Control Information}
\newacronym{dl}{DL}{Downlink}
\newacronym{dmr}{DMR}{Deadline Miss Ratio}
\newacronym{dmrs}{DMRS}{DeModulation Reference Signal}
\newacronym{dray}{D-Ray}{Deterministic Ray}
\newacronym{e2e}{E2E}{end-to-end}
\newacronym{ecn}{ECN}{Explicit Congestion Notification}
\newacronym{edf}{EDF}{Earliest Deadline First}
\newacronym{em}{EM}{electromagnetic}
\newacronym{enb}{eNB}{evolved Node Base}
\newacronym{endc}{EN-DC}{E-UTRAN-\gls{nr} \gls{dc}}
\newacronym{epc}{EPC}{Evolved Packet Core}
\newacronym{es}{ES}{Edge Server}
\newacronym{fdd}{FDD}{Frequency Division Duplexing}
\newacronym{fdma}{FDMA}{Frequency Division Multiple Access}
\newacronym{fray}{F-Ray}{Flashing Ray}
\newacronym{fs}{FS}{Fast Switching}
\newacronym{ftp}{FTP}{File Transfer Protocol}
\newacronym{gmm}{GMM}{Gaussian Mixture Model}
\newacronym{gnb}{gNB}{Next Generation Node Base}
\newacronym{harq}{HARQ}{Hybrid Automatic Repeat reQuest}
\newacronym{hetnet}{HetNet}{Heterogeneous Network}
\newacronym{hh}{HH}{Hard Handover}
\newacronym{hol}{HOL}{Head-of-Line}
\newacronym{hqf}{HQF}{Highest-quality-first}
\newacronym{ia}{IA}{Initial Access}
\newacronym{iab}{IAB}{Integrated Access and Backhaul}
\newacronym{imt}{IMT}{International Mobile Telecommunication}
\newacronym{inf}{InF}{Indoor Factory}
\newacronym{irs}{IRS}{Intelligent Reflective Surface}
\newacronym{inr}{INR}{Interference to Noise Ratio}
\newacronym{iot}{IoT}{Internet of things}
\newacronym{ked}{KED}{Knife-Edge Diffraction}
\newacronym{kpi}{KPI}{Key Performance Indicator}
\newacronym{lcf}{LCF}{Level Crossing Frequency}
\newacronym{lcr}{LCR}{Level Crossing Rate}
\newacronym{los}{LoS}{Line-of-Sight}
\newacronym{l2sm}{L2SM}{Link-to-System Mapping}
\newacronym{lte}{LTE}{Long Term Evolution}
\newacronym{ltemtp}{LTE-M}{LTE-MTC [Machine Type Communication]}
\newacronym{m2m}{M2M}{Machine to Machine}
\newacronym{mac}{MAC}{Medium Access Control}
\newacronym{mc}{MC}{Markov chain}
\newacronym{mcs}{MCS}{Modulation and Coding Scheme}
\newacronym{mdp}{MDP}{Markov decision process}
\newacronym{mec}{MEC}{Mobile Edge Cloud}
\newacronym{mi}{MI}{Mutual Information}
\newacronym{mib}{MIB}{Master Information Block}
\newacronym{mimo}{MIMO}{Multiple Input Multiple Output}
\newacronym{m-mimo}{m-MIMO}{massive-MIMO}
\newacronym{mlr}{MLR}{Maximum-local-rate}
\newacronym{mmwave}{mmWave}{millimeter wave}
\newacronym{moi}{MoI}{Method of Images}
\newacronym{mpc}{MPC}{Multi Path Component}
\newacronym{mptcp}{MPTCP}{Multipath TCP}
\newacronym{mr}{MR}{Maximum Rate}
\newacronym{mrdc}{MR-DC}{Multi \gls{rat} \gls{dc}}
\newacronym{mss}{MSS}{Maximum Segment Size}
\newacronym{mtd}{MTD}{Machine-Type Device}
\newacronym{mtu}{MTU}{Maximum Transmission Unit}
\newacronym{ne}{NE}{Nash equilibrium}
\newacronym{nfv}{NFV}{Network Function Virtualization}
\newacronym{nist}{NIST}{National Institute of Standards and Technology}
\newacronym{nlos}{NLoS}{Non-Line-of-Sight}
\newacronym{ntn}{NTN}{Non-Terrestrial Networks}
\newacronym{nr}{NR}{New Radio}
\newacronym{nrmse}{NRMSE}{Normalized Root Mean Square Error}
\newacronym{nsa}{NSA}{Non Stand Alone}
\newacronym{o2i}{O2I}{Outdoor-to-Indoor}
\newacronym{ofdm}{OFDM}{Orthogonal Frequency Division Multiplexing}
\newacronym{pa}{PA}{power amplifier}
\newacronym{prr}{PRR}{Packet Reception Ratio}
\newacronym{pbch}{PBCH}{Physical Broadcast Channel}
\newacronym{pdcch}{PDCCH}{Physical Downlonk Control Channel}
\newacronym{pdcp}{PDCP}{Packet Data Convergence Protocol}
\newacronym{pdsch}{PDSCH}{Physical Downlink Shared Channel}
\newacronym{pdu}{PDU}{Packet Data Unit}
\newacronym{per}{PER}{Packet Error Rate}
\newacronym{pf}{PF}{Proportional Fair}
\newacronym{pgw}{PGW}{Packet Gateway}
\newacronym{phy}{PHY}{Physical}
\newacronym{pl}{PL}{Path Loss}
\newacronym{PoA}{PoA}{price of anarchy}
\newacronym{ppp}{PPP}{Poisson Point Process}
\newacronym{prb}{PRB}{Physical Resource Block}
\newacronym{psd}{PSD}{Power Spectral Density}
\newacronym{pss}{PSS}{Primary Synchronization Signal}
\newacronym{pucch}{PUCCH}{Physical Uplink Control Channel}
\newacronym{pusch}{PUSCH}{Physical Uplink Shared Channel}
\newacronym{qam}{QAM}{Quadrature Amplitude Modulation}
\newacronym{qd}{QD}{Quasi Deterministic}
\newacronym{rach}{RACH}{Random Access Channel}
\newacronym{ran}{RAN}{Radio Access Network}
\newacronym[firstplural=Radio Access Technologies (RATs)]{rat}{RAT}{Radio Access Technology}
\newacronym{red}{RED}{Random Early Detection}
\newacronym{rf}{RF}{Radio Frequency}
\newacronym{rlc}{RLC}{Radio Link Control}
\newacronym{rlf}{RLF}{Radio Link Failure}
\newacronym{rray}{R-Ray}{Random Ray}
\newacronym{rrc}{RRC}{Radio Resource Control}
\newacronym{rrm}{RRM}{Radio Resource Management}
\newacronym{rs}{RS}{Remote Server}
\newacronym{rsrp}{RSRP}{Reference Signal Received Power}
\newacronym{rsrq}{RSRQ}{Reference Signal Received Quality}
\newacronym{rss}{RSS}{Received Signal Strength}
\newacronym{rssi}{RSSI}{Received Signal Strength Indicator}
\newacronym{rt}{RT}{Ray Tracer}
\newacronym{rtt}{RTT}{Round Trip Time}
\newacronym{rw}{RW}{Receive Window}
\newacronym{rx}{RX}{Receiver}
\newacronym{sa}{SA}{standalone}
\newacronym{sack}{SACK}{Selective Acknowledgment}
\newacronym{sap}{SAP}{Service Access Point}
\newacronym{sch}{SCH}{Secondary Cell Handover}
\newacronym{scm}{SCM}{Spatial Channel Model}
\newacronym{scoot}{SCOOT}{Split Cycle Offset Optimization Technique}
\newacronym{sdap}{SDAP}{Service Data Adaptation Protocol}
\newacronym{sdma}{SDMA}{Spatial Division Multiple Access}
\newacronym{sf}{SF}{Shadow Fading}
\newacronym{si}{SI}{Study Item}
\newacronym{sib}{SIB}{Secondary Information Block}
\newacronym{sinr}{SINR}{Signal-to-Interference-plus-Noise Ratio}
\newacronym{sir}{SIR}{Signal-to-Interference Ratio}
\newacronym{sm}{SM}{Saturation Mode}
\newacronym{snr}{SNR}{Signal-to-Noise Ratio}
\newacronym{son}{SON}{Self-Organizing Network}
\newacronym{srs}{SRS}{Sounding Reference Signal}
\newacronym{ss}{SS}{Synchronization Signal}
\newacronym{sss}{SSS}{Secondary Synchronization Signal}
\newacronym{sta}{STA}{Station}
\newacronym{tb}{TB}{Transport Block}
\newacronym{tcp}{TCP}{Transmission Control Protocol}
\newacronym{udp}{UDP}{User Datagram Protocol}
\newacronym{tdd}{TDD}{Time Division Duplexing}
\newacronym{tdma}{TDMA}{Time Division Multiple Access}
\newacronym{tfl}{TfL}{Transport for London}
\newacronym{tgad}{TGad}{Task Group ad}
\newacronym{tgay}{TGay}{Task Group ay}
\newacronym{tm}{TM}{Transparent Mode}
\newacronym{trp}{TRP}{Transmitter Receiver Pair}
\newacronym{tti}{TTI}{Transmission Time Interval}
\newacronym{ttt}{TTT}{Time-to-Trigger}
\newacronym{tx}{TX}{Transmitter}
\newacronym{ue}{UE}{User Equipment}
\newacronym{ul}{UL}{Uplink}
\newacronym{um}{UM}{Unacknowledged Mode}
\newacronym{uma}{UMa}{Urban Macro}
\newacronym{uml}{UML}{Unified Modeling Language}
\newacronym{utc}{UTC}{Urban Traffic Control}
\newacronym{vm}{VM}{Virtual Machine}
\newacronym{wbf}{WBF}{Wired Bias Function}
\newacronym{wf}{WF}{Wired-first}
\newacronym{wifi}{Wi-Fi}{Wireless Fidelity}
\newacronym{wigig}{WiGig}{Wireless Gigabit}
\newacronym{wlan}{WLAN}{Wireless Local Area Network}
\newacronym{xpr}{XPR}{Cross Polarization Ratio}
\newacronym{fr2}{FR2}{Frequency Range 2}
\newacronym{nbiot}{NB-IoT}{Narrowband-IoT}
\newacronym{cps}{CPS}{Cyber-Physical production System}
\newacronym{iiot}{IIoT}{Industrial Internet of things}
\newacronym{agv}{AGV}{Autonomous Ground Vehicle}
\newacronym{uav}{UAV}{Unmanned Autonomous Vehicle}
\newacronym{amr}{AMR}{Autonomous Mobile Robots}
\newacronym{wsn}{WSN}{Wireless Sensor Network}
\newacronym{embb}{eMBB}{enhanced Mobile Broadband}
\newacronym{urllc}{URLLC}{Ultra-Reliable Low-Latency Communications}
\newacronym{vi}{VI}{value iteration}

\title{A Markov Game of Age of Information From Strategic Sources With Full Online Information}

\author{
    \IEEEauthorblockN{Matteo~Pagin, Leonardo~Badia, and Michele~Zorzi}
    \IEEEauthorblockA{Department of Information Engineering (DEI), University of Padova, Italy \\
    \{paginmatte,badia,zorzi\}@dei.unipd.it
    }
    \thanks{This work was supported by the RESTART Program, financed by the Italian government with the resources of the PNRR -- Mission 4, Component 2, Investment 1.3, theme 14 ``Telecommunications of the future.'' }
}

\maketitle

\begin{abstract}
We investigate the performance of concurrent remote sensing from independent strategic sources, whose goal is to minimize a linear combination of the freshness of information and the updating cost.
In the literature, this is often investigated from a static perspective of setting the update rate of the sources a priori, either in a centralized optimal way or with a distributed game-theoretic approach. However, we argue that truly rational sources would better make such a decision with full awareness of the current age of information, resulting in a more efficient implementation of the updating policies. To this end, we investigate the scenario where sources independently perform a stateful optimization of their objective. Their strategic character leads to the formalization of this problem as a Markov game, for which we find the resulting Nash equilibrium. This can be translated into practical smooth threshold policies for their update. The results are eventually tested in a sample scenario, comparing a centralized optimal approach with two distributed approaches with different objectives for the players.
\end{abstract}

\begin{picture}(0,0)(0,-350)
\put(0,0){
\put(0,0){\qquad \qquad \quad This paper has been submitted to IEEE for publication. Copyright may change without notice.}}
\end{picture}


\section{Introduction}
Pervasive real-time sensing is a key component of several use cases for future generation wireless networks, such as multisensory communication for digital twins, \gls{arvr}, robots, eHealth, Industry 4.0 and so on \cite{viswanathan2020communications,giordani2020toward}. Sensing capabilities are often not delegated to a single terminal but distributed across the network among multiple devices and even more numerous logical entities.

In such a context, freshness of information becomes very relevant and can be characterized by the key performance indicator of the \gls{aoi} \cite{kaul2012real,munari}. For these aforementioned use cases, having up-to-date information is possibly more interesting than optimizing raw throughput or average latency, and can be often precisely quantified through closed-form expressions, which makes it appealing for analytical investigations.

At the same time, to obtain scalability and ease of implementation across pervasive multi-terminal scenarios such as the Internet of Things, a distributed management with low complexity algorithms is required \cite{delsing2017local,yang2021understanding}. This motivates a game theoretic approach to the evaluation of AoI coming from multiple sources of information \cite{yates2012real,badia2021game,saurav2021game}.

In light of these motivations, in the present paper we consider a scenario consisting of independent sources, tracking the same process of interest for an information sink, to which they may independently send periodic updates at the price of incurring an individual cost \cite{hribar2018using,badia2021age}.
We also assume that all sources can identically send a valuable information update that refreshes the AoI at the sink node. At the same time, they are also aware that sending redundant updates, when another source is already doing so, causes extra costs and does not further lower the AoI beyond the level already reached.

If we frame the strategic decision-making mechanism of the devices through game theory, we obtain that they are facing a classic dilemma. On the one hand, sending an update implies a cost that each source would prefer not to pay, as the AoI can be identically decreased if another source sends an update (but there is no cost to pay in this case). However, if all nodes decide not to update, they end up causing the AoI to soar, which is not efficient either.

To solve this conundrum, we frame the problem as a Markov game, on which we perform a distributed optimization from the individual perspective resulting in a \gls{ne} \cite{aras2004cooperation,leng2019age}. We can actually consider the strategic objective for a single node to be a selfish optimization of the \gls{aoi} under individual cost, or the overall minimization involving the update cost for the entire network, but still from a local perspective. Finally, we can take the global network optimization achieved by a centralized approach as a benchmark.

Our analysis is original as only a handful of references study a distributed strategic optimization of AoI, and they all assume either \emph{competing} terminals as in \cite{saurav2021game} or a stateless optimization \cite{badia2021age}, while we are the first to consider a stateful minimization of a common AoI objective.

Our results indicate that, compared with an optimal centralized approach, distributing the choice of updating causes a smoothing of the individual transmission probabilities. Some threshold effects are still present: the nodes do not update when the information is fresh, but also may fail to update with certainty when the information is stale. This is due to the lack of coordination, i.e., nodes may be strategically transferring the burden of updating to one another, to avoid paying the cost, resulting in a high \gls{PoA} \cite{prospero2021resource}.

As a consequence, a distributed selfish system achieves worse performance with full online information than %
when the sources are not aware of the system AoI,
%what would be achieved by the same users but working without knowing the actual \gls{aoi} of the sink, 
in which case a close-to-optimal performance is achieved \cite{badia2021game}. This can be seen as a consequence of providing strategic players with more information, which can be harmful in selfish setups \cite{li2019stackelberg,canzian2012promoting}. 
Nevertheless, our results show practical consequences that ought to be taken into account in the network design of remote sensing protocols, possibly designing effective rewarding mechanisms to benefit the strategic users \cite{hu2020blockchain}.

The rest of this paper is organized as follows. In Section \ref{sec:mod}, we characterize the system model to develop our game theoretic analysis. Section \ref{sec:theo} proves some theoretical findings that are further quantitatively visualized in Section \ref{sec:res} through numerical results. Finally, Section \ref{sec:concs} gives the conclusions.

\section{System model}
\label{sec:mod}

%\todo{Would it be better to split this into different sections ? I find it difficult to link the various paragraphs of the current system model}

We consider a system where $N$ sources $S_1, \ldots ,S_N$ transmit updates related to a common underlying process to a receiving gateway $R$. We measure the value of such updates as the ``freshness" of its information, formally defined as the \gls{aoi}, i.e., the difference between the current time and the instant of the last update~\cite{kaul2012real}. 
In our statement, these sources are not competing with one another nor actively collaborating. Instead, they independently send updates to $R$, being aware that receiving an update at the same time instant from multiple sources results in wasting resources. To account for this, we consider that each transmission incurs a cost $c$ \cite{gindullina2021age}. Indeed, without such a term, the sources would be transmitting at every possible instant, making the problem at hand trivial. Moreover, the model would not be consistent with the physical world, where sensors exhibit energy and processing limitations, especially in \gls{iot} scenarios \cite{hribar2018using}. 

In line with similar analyses, we sample the system at periodic instants, obtaining a discrete-time axis of possible update epochs. Accordingly, \gls{aoi} $\delta$ at one such instant $t$ reads:
\vspace{-0.4cm}
\begin{equation}
\delta (t) = t - \textrm{max} \bigg( \big( \{ \tau \leq t \} \big) \cap \Big( \bigcup^N_{i=0} \bm{\tau}_i \Big)\bigg),
\end{equation}
 where the set $\bm{\tau}_i$ collects the time instants of the $i$-th source updates, i.e., $\bm{\tau}_i \doteq \{ \ldots, \tau_i^{(1)}, \tau_i^{(2)}, \ldots, \tau_i^{(n)}, \ldots \}$.
In each time slot, the sources independently decide whether to transmit an update with probability $p_i$ that depends on the current value of the \gls{aoi}, i.e., $p_i \doteq \mathbb{P} \left[ \text{transmitting} \, | \, \delta = i \right]$. The knowledge of the \gls{aoi} at the sources can be justified by assuming either that $R$ sends a broadcast acknowledgment whenever an update is received, or that the sources eavesdrop the channel in a \gls{csma} fashion \cite{tourrilhes1998robust}. Then, this information is used to keep track of the evolution of the \gls{aoi} over time\footnote{The value of $N$ can also be estimated in a distributed manner, for instance using the approach of~\cite{pang2014network}.}. 
Finally, we assume that updates are always correctly received at $R$, since accounting for the presence of errors is already done in many papers~\cite{yates2017status}, and
including it here would distract the reader from the main focus of this analysis, i.e., quantifying the loss in efficiency caused by a lack of coordination.

We model the problem of finding the optimal transmission probabilities as a static game of complete information $\mathcal{G} = (\mathcal{S}, \mathcal{A}, \mathcal{R})$. The set of players $\mathcal{S}=\{ S_1, \ldots, S_N \}$ includes the sources, as the gateway is only a passive receiver of updates, incapable of making any move, hence it is not part of $\mathcal{S}$. 
The action set $\mathcal{A}$ contains the possible update probabilities $p_i$'s, equal for each source. In $\mathcal{G}$, the players choose such transmission probabilities in one shot, based on the expected discounted reward over an infinite time horizon. Thus, we model the evolution of the \gls{aoi} as a discrete-time, countably infinite \gls{mc} where the Markov property is satisfied by definition of $p_i$ and the state-space includes all possible values of the \gls{aoi}, i.e., $\mathbb{Z}_+$. From a generic state $i$, only states $i + 1$ and $0$ are accessible in one step, as the \gls{aoi} either increases by one if no transmission occurs, or goes to zero whenever at least one of the sources transmits. The transition probabilities of the MC are
\begin{equation}
p_{i, j} = \begin{cases}
(1 - p_i)^N \;\; \forall \, i \in \mathbb{Z}_+ & j = i + 1\\
1 - (1 - p_i)^N \;\; \forall \, i \in \mathbb{Z}_+ & j = 0\\
0 &\text{ otherwise.}
\end{cases}
\end{equation} 

Finally, we define the reward $\mathcal{R}$ considering the \textit{unilateral} payoff of each source instead of the system overall reward \cite{yang2021understanding}.
We first define the overall expenditure $K$ of a source as the sum of the \textit{system} \gls{aoi} and the \textit{individual} transmission cost. Then, we define the expected \textit{distributed selfish} reward as $\mathbb{E} \left[ \mathcal{R}^s (i, p_i) \right]$ from state $i$ and given action $p_i$ as:
\begin{equation}
\begin{split}
\mathbb{E} \left[ \mathcal{R}^s (i, p_i) \right] & = \mathbb{E} \left[ \mathcal{R}_{t + 1}^s \, | \, \mathcal{S}_t = i, \mathcal{A}_t = p_i \right]   \\
& = -\mathbb{E} \left[ K (i, p_i) \right] \\
& = - (i+1) (1 - p_i)^N - c  p_i.
\end{split}
\end{equation} 
%Likewise, we define the distributed selfish policy as the set of actions $\{p_i\}_i$ which maximizes the corresponding reward function. 

We also consider a \textit{centralized} policy, which implies that the nodes coordinate, with the goal of avoiding multiple concurrent transmissions. This corresponds to an equivalent system with just one source, whose expected reward thus reads
\begin{equation}
\begin{split}
\mathbb{E} \left[ \mathcal{R}^c (i, p_i) \right] & = \mathbb{E} \left[ \mathcal{R}_{t + 1}^c \, | \, \mathcal{S}_t = i, \mathcal{A}_t = p_i \right]   \\
& = - (i+1) (1 - p_i) - c p_i \, \forall \, N \in \mathbb{Z}_+ .
\end{split}
\end{equation}  

Finally, we define a \textit{distributed global} policy that is somehow intermediate, in that the costs incurred by the system are locally computed and all the nodes share the objective of minimizing the network AoI. 
In this way, we take the existence of multiple nodes tracking the same process into account, while still considering a distributed approach. In this case, the expected reward reads:
\begin{equation}
\begin{split}
\label{eq:reward_global}
\mathbb{E} \left[ \mathcal{R}^g (i, p_i) \right] & = \mathbb{E} \left[ \mathcal{R}_{t + 1}^g \, | \, \mathcal{S}_t = i, \mathcal{A}_t = p_i \right]   \\
& = - (i+1) (1 - p_i)^N - N \cdot c p_i .
\end{split}
\end{equation} 
The two distributed policies consider different approaches to decentralized management; in \textit{distributed selfish}, all the nodes are anarchical~\cite{prospero2021resource}, whereas in \textit{distributed global} they have the same goal but act without coordination, which may still decrease the efficiency from a centralized control~\cite{badia2021age}.

We define a \gls{mdp} $(\hat{\mathcal{S}}, \hat{\mathcal{A}}, \hat{\mathcal{P}}, \hat{\mathcal{R}})$ on top of the \gls{mc}, with the aim of computing the set(s) of transmission probabilities leading to \glspl{ne}. The set of transition probabilities $\hat{\mathcal{P}}$ and the state-space $\hat{\mathcal{S}}$ coincide with their counterparts of the \gls{mc}. The action space $\hat{\mathcal{A}}$ is represented by the $p_i$'s, equal for each source for symmetry reasons, and stationary. For practical purposes, we discretize the possible transmission probabilities into $k$ values as $\hat{\mathcal{A}} \doteq \{0, \Delta, \ldots, (k - 1) \Delta\} \, ; \, \Delta = 1/(k-1)$. For similar reasons, we bound the state space as $\hat{\mathcal{S}} = \{ n \in  \mathbb{Z}_+ \, | \, n < \delta_{max}\} $.

From the \gls{mdp} parameters, we find the optimal policy by using \gls{vi}~\cite{sutton2018reinforcement}. We introduce a discount factor $\gamma \,{\in}\, \left[ 0, 1 \right[$ to obtain a finite expected return despite the infinite horizon. Then, we estimate the discounted value-state function $ v^n (s, \pi) \doteq \mathbb{E} \left[ \sum_{k = 0}^{+ \infty} \gamma^k \mathcal{R}_{t + k + 1}^n \, | \, \mathcal{S}_t = s, \, \pi = \left\{ p_1^{\pi}, \ldots, p_{\delta{max}}^{\pi} \right\} \right] \, \forall \, s {\in} \hat{\mathcal{S}}, n \in \{s, g, c\}$ by repeating the following update for $K$ iterations:
\begin{equation}
\begin{split}
v_{k + 1}^n (s, \pi) & \doteq \underset{p_i}{\mathrm{max}} \, \mathbb{E} \left[ \mathcal{R}_{t + 1}^n + \gamma v_k^n (\mathcal{S}_{t + 1}, \pi) | \, \mathcal{S}_t = i, \mathcal{A}_t = p_i^{\pi} \right]   \\
& = \underset{p_i}{\mathrm{max}} \, \left[ \mathcal{R}^n(i, p_i^{\pi}) + \sum_j \gamma \, p(j \, | \, i, p_i^{\pi}) v_k^n (j, \pi)  \right]
\end{split}
\nonumber
\end{equation} 
and obtaining the optimal policy $\pi^*$, where the transmission probability for the $i$-th state $p_i^{\pi^*}$ satisfies a pseudo-steady-state condition $\underset{p_i}{\mathrm{argmax}} \, v^{n}_K (i, \pi) \approx \underset{p_i}{\mathrm{argmax}} \, v^{n} (i, \pi) $ .

\section{Theoretical Analysis} %change name
\label{sec:theo}
%\textcolor{red}{LEO: this is my tentative proof part. It is to be rearranged, of course, but I wanted to see how it looks in LaTeX.}

We now prove some structural results on the activation threshold depending on the state of the AoI~\cite{gindullina2021age}.

Define $\bar{K}(i, \pi) \doteq - v^{s}(i, \pi) $ as the expected long-term total discounted cost starting from state $i$, choosing policy $\pi$. Let also $p_{i}^* \doteq  p_i^{\pi^*}$, for the sake of readability. We prove Theorems \ref{thm:1} and \ref{lemma:1} for the case of a single terminal sending data updates, and then generalizing to multiple transmitters.

%\subsection*{Single transmitter.}  

\begin{thm}
\label{thm:1}
For $\gamma \to 1$, the optimal policy $\pi^*$ satisfies either $p_i^* = 1$ or $p_i^*=0$, depending on the specific state $i$ and following a threshold behavior, i.e., $p_i^* = 1$ if and only if $i$ is greater than or equal to a threshold value, which depends on the cost (the larger the cost, the higher the threshold).
\end{thm}

\begin{IEEEproof} 
At state $i$, the expected long-term discounted cost is the sum of three terms, namely: the transmission cost, which is equal to $c p_i$ since the node transmits with probability $p_i$; the expected discounted cost from state $0$, which is considered if the transmission is performed and therefore is equal to $\gamma p_i \bar{K}(0,\pi)$, and finally the expected discounted cost from state $i+1$, which is considered if the transmission is not performed instead, i.e., the related term is $\gamma (1-p_i) \bar{K}(i +1 + (i+1, \pi))$. Thus, 
\begin{equation}
\begin{aligned}
\bar{K}(i,\pi) = \, & c p_i + \gamma p_i \bar{K}(0,\pi) + \\
& \gamma (1-p_i) (i+1 + \bar{K}(i+1, \pi)),
\end{aligned}
\label{eq:single_terminal}
\end{equation}

\noindent that, in turn, can be re-arranged to obtain

\begin{equation}
p_i = \frac{ \gamma \left[ i+1+ \bar{K}(i+1, \pi) \right]  - \bar{K}(i, \pi) }{ \gamma \left[ i+1 + \bar{K}(i+1, \pi) - \bar{K}(0, \pi) \right] - c },
\end{equation}

\noindent from which it is easy to conclude that: (a) the $p_i^*$ minimizing the average cost $\bar{K}(i, \pi^*)$ is non-decreasing in $i$, due to the only coefficient of $p_i$ in (\ref{eq:single_terminal}) that depends on $i$ being $-(i{+}1)$; and (b) the limit value of $p_i^*$ for $i\to +\infty$ is $1$.

Conversely, (\ref{eq:single_terminal}) can also be exploited to say that, for a fixed value of $i$, the choice of $p_i$ results in a monotonic behavior depending on the coefficients in front of the $p_i$ terms. Those with a negative sign, i.e., decreasing the cost as $p_i$ increases are $i+1+\bar{K}(i+1, \pi)$, whereas those with a positive sign are $\bar{K}(0, \pi)$ and the cost $c$. Thus, it follows that if the cost is $0$, the highest possible value of $p_i$ will be chosen, i.e., $p_i^* = 1$. The same will happen if $c$ is small, i.e., not able to make the sum of the positive coefficients greater than that of the negative ones. When this happens, the cost-minimizing value of $p_i^*$ will be $0$. For a different $i$, the trend is still the same but according to the previous reasoning a larger $c$ is required to activate the transmissions.
\end{IEEEproof}

\begin{thm}
\label{lemma:1}
For $\gamma \to 1$, states $s_1$ and $s_2>s_1$:
\begin{equation}
\bar{K}(s_2, \pi) - \bar{K}(s_1, \pi) \leq s_2 - s_1.
\end{equation}
\end{thm}

\begin{IEEEproof} 
Let $\mathcal{A}$ be the class of the \gls{mdp} underlying \gls{mc} states such that their transmission probability $p_i$ is strictly less than $1$. It follows from Theorem~\ref{thm:1} that this class is of the form $\mathcal{A} = \{n \in  \mathbb{Z}_+ \, | \, n < \vartheta \leq \delta_{max}\}$. Hence, it is irreducible, since all of its states communicate with each other:
\begin{equation}
p_{s_1, s_2}^{s_2 - s_1} \geq p_{s_1, s_1+1} \cdot \ldots \cdot p_{s_2 - 1, s_2} > 0
\end{equation}
\vspace*{-0.3cm}
and
\vspace*{-0.1cm}
\begin{equation}
p_{s_2, s_1}^{s_1 + 1} \geq p_{s_2, 0} \cdot p_{0, 1} \cdot \ldots \cdot p_{s_1 - 1, s_1} > 0 ,
\end{equation}
where $p_{s_1, s_2}^n$ is the probability of reaching state $s_2$ starting from state $s_1$ in exactly $n$ steps.
Thus, all states belonging to $\mathcal{A}$ are eventually reached in a finite number of steps and the \gls{mdp} will collect, during its transition from $s_1$ to $s_2$, a finite reward. 
From that state onward, the evolution of the \gls{mc} is statistically equivalent to that obtained when starting from $s_2$ due to the Markov property. It follows that
\begin{equation}
\lim_{\gamma \to 1} \bar{K}(s_2, \pi) - \bar{K}(s_1, \pi) = 0 .
\end{equation}
Then, the result follows directly from the definition of limit in the $(\epsilon, \delta)$-sense, taking $\epsilon = s_2 - s_1$.
\end{IEEEproof} 

These findings state that the single transmitter never updates the information until the AoI is large enough, then it will transmit with probability $1$ once a threshold state $i = \vartheta$ is reached. This means that all states $i>\vartheta$ will never be reached and the transmitter will count from $0$ to $\vartheta$. Thus, the long-term average AoI will be $\vartheta/2$ and the long-term average transmission cost will be $c/(\vartheta+1)$ since updates will happen periodically every $\vartheta+1$ slots.
Such $\vartheta$ must be equal to $\lceil \sqrt{2c} - 1 \rceil$, which means that a cost coefficient $c$ below $0.5$ will be ineffective in limiting the transmissions~\cite{badia2021age}.

For multiple sources, a similar reasoning to Theorems \ref{thm:1} and \ref{lemma:1} can be applied.
\begin{comment}
{{\color{blue}{First, the optimal policy must lead to a recurrent \gls{mc}, i.e., it must transmit at some point in time. Indeed, the policy that never transmits, i.e., $p_i=0$ for all $i$, achieves an unbounded AoI. %However, the long-term expected discounted cost is still limited, because of the discount factor (precisely, the achieved value of $\bar{K}(0,\pi)=\sum_{j} j\gamma^j = \gamma/(1-\gamma)^2$.
Hence, this policy cannot be optimal, as it is surely worse than transmitting if and only if the AoI is above $cN$.
%(this is likely not optimal either, but it just serves to show that the optimal policy must be recurrent.
}}
\end{comment}
It is convenient to use $\sigma_i \doteq 1-p_i$ as the probability of a terminal being silent in state $i$, and to define the auxiliary variable $y_i \doteq \sigma_i^N$. %Consequently, we will write the expected long-term cost at state $i$ as $\bar{K}(i, \pi)$.

\oldversion{
If we proceed along the same lines of~(\ref{eq:single_terminal}), we can write
\begin{IEEEeqnarray}{rCl}
\bar{K}(i,\pi) &=& c (1-\sigma_i) + \gamma \bar{K}(0,\pi) (1-\sigma_i^N)  \\
&&+ \gamma \sigma_i^N \big( i+1+\bar{K}(i+1,\pi) \big) \nonumber
\end{IEEEeqnarray}
and, approximating $i$ to a continuous variable, take the first-order derivative of this equation to get
\begin{IEEEeqnarray}{rCl}
\frac{\partial \bar{K}(i,\pi)}{\partial i} &=& - c \frac{\partial \sigma_i}{\partial i}
+ \gamma N \big( i+1+\bar{K}(i+1,\pi) \nonumber \\
&& - \bar{K}(0,\pi) \big) \sigma_i^{N{-}1} \frac{\partial \sigma_i}{\partial i} + \gamma \sigma_i^{N} \,.
\label{eq:multiple_terminals}
\end{IEEEeqnarray}

While (\ref{eq:multiple_terminals}) contains more involved terms than (\ref{eq:single_terminal}), it can be used to derive similar, albeit less clear-cut, conclusions.
In particular, we can remark that $\bar{K}(i+1,\pi) - \bar{K}(0,\pi)$ is always less than or equal to $i+1$ (being exactly identical if all terminals are silent in states $0,1, \dots, i$) and therefore the total cost is increasing in $\sigma_i$ since the differential terms of $\bar{K}(i,\pi)$ and $\sigma_i$ all have the same sign except for the transmission cost, which is negative. This means that similar conclusions to the previous case can be drawn as (a) under the optimal policy, the transmission probability increases (i.e., $\sigma_i$ decreases) in $i$, and (b) it ultimately reaches $1$ when $i$ goes to infinity.
}

If we proceed along the same lines of~(\ref{eq:single_terminal}), we can write
\begin{IEEEeqnarray}{rCl}
\label{eq:multiple_terminals}
\bar{K}(i,\pi) &=& c (1-\sigma_i) + \gamma \bar{K}(0,\pi) (1- y_i)  \\
&& + \, \gamma y_i \big( i+1+\bar{K}(i+1,\pi) \big), \nonumber 
\end{IEEEeqnarray}
which can be manipulated into
\begin{equation}
\label{eq:multiple_terminals_line}
y_i = \frac{\bar{K}(i,\pi) - \gamma \bar{K}(0,\pi) - (1 - \sigma_i)c}{ \gamma\left[ \bar{K}(i + 1,\pi) + i + 1 - \bar{K}(0,\pi) \right]} 
% +\gamma y_i \big( i+1+\bar{K}(i+1,\pi) \big), \nonumber 
\end{equation}
and $\sigma_i$ is simultaneously satisfying $y_i = \sigma_i^N$ and
(\ref{eq:multiple_terminals_line}).

While these equations contain more involved terms than (\ref{eq:single_terminal}), they can be used to derive similar, albeit less clear-cut, conclusions.
From (\ref{eq:multiple_terminals}), we can remark that the expected long-term discounted cost at state $i$ is increasing in $\sigma_i$ and similar conclusions to the previous case can be drawn from (\ref{eq:multiple_terminals_line}
) as (a) under the optimal policy, the transmission probability increases (i.e., $\sigma_i$ decreases) in $i$, and (b) it ultimately tends to $1$ when $i$ goes to infinity.

However, due to the exponent $N$ that was not present in the single-terminal case, the analogies are limited to the monotonic character, whereas the binary behavior of the terminal being always active or inactive no longer applies. On the contrary, instead of the terminal becoming active once a sufficiently high value of the AoI is reached, (\ref{eq:multiple_terminals}) implies a smoother increasing behavior of the transmission probability versus the AoI of the system that goes to $1$ only asymptotically.

Still, if $c$ is very large and the value of $i$ is not high enough, the minimization of the total cost would require a negative value for $p_i$, which is not admissible. Thus, we can prove that the behavior of the transmission probability $p_i$ for increasing $i$ in a multi-terminal case is non-decreasing, possibly starting from zero for low values of $i$, then gradually increases, and only tends to $1$ in the limit for very high values of $i$.

%\begin{footnotesize}
%\[
%\begin{aligned}
%&v_{\pi}(0)=E[R^s(0, \pi(0))]+\sum_{s'} p_{0, s'} v_{\pi}\left(s'\right) \\
%&v_{\pi}(0)=-\left(1-p_{0}\right)^{N}-c p_{0}+p_{0,0} \cdot v_{\pi}(0)+p_{0,1} \cdot v_{\pi}(1) \\
%&v_{\pi}(0)=-\left(1-p_{0}\right)^{N}-cp_{0}+\left[1-\left(1-p_{0}\right)^{N}\right] v_{\pi}(0)+ \left(1-p_{0}\right)^{N} v_{\pi}(1) \\
%&v_{\pi}(0)\left[1-\left(1-\left(1-p_{0}\right)^{N}\right)\right]=-\left(1-p_{0}\right)^{N}-cp_{0}+\left(1-p_{0}\right)^{N} \cdot v_{\pi}(1) \\
%&v_{\pi}(0)\left(1-p_{0}\right)^{N}=-\left(1-p_{0}\right)^{N}-c p_{0}+\left(1-p_{0}\right)^{N} \cdot v_{\pi}(1) \\
%&\left(1-p_{0}\right)^{N}\left[ v_{\pi}(0)-v_{\pi}(1) \right]= - (1-p_0)^N -c p_{0}  \\
%&v_{\pi}(0)-v_{\pi}(1)= -1 -\frac{c p_{0}}{\left(1-p_{0}\right)^{N}} < 0
%\end{aligned}
%\]
%\end{footnotesize}

\section{Numerical evaluations}
\label{sec:res}

%\todo{Here I am adding all the figures which I make, then we can choose a subset of them to be actually kept in the paper. I am also adding a few comments on the figures which I believe we will keep.}

We numerically evaluate the policies that arise from the different system configurations, with the goal of validating the theoretical results presented in the previous section. %which in classic game-theoretic fashion models the players as rational and selfish entities.
First, we consider the update probabilities $u_i$, defined as the probability that at least one source is transmitting in a given state $i$, that is, $u_i = 1 - \sigma_i^N = 1 - (1-p_i^*)^N$. Then, we compute the stationary probabilities $\pi_i$ of the underlying \gls{mc}, given the optimal policy, and we inspect the average behavior of the system in terms of resulting average update probability $\bar{U}$, \gls{aoi} $\bar{\Delta}$, system cost $\bar{C}$ and reward $\bar{\mathcal{R}}$, defined as

\begin{equation*}
\bar{U} = \sum_{i}  \pi_i u_i
\qquad
\bar{\Delta} = \sum_{i}  \pi_i i
\end{equation*} 
\vspace{-3mm}
\begin{equation*}
\bar{C} = \sum_{i}  - \pi_i \left[ 1 + (N-1) \cdot \mathbbm{D} \right] \cdot c p_i^{*}
\end{equation*} 
\vspace{-3mm}
\begin{equation*}
\bar{\mathcal{R}} = \sum_{i}  -\pi_i (i + 1)(1 - p_i^{*})^{1 + (N-1) \cdot \mathbbm{D}} - \bar{C},
\end{equation*} 
where $ \mathbbm{D}$ is the indicator function which maps to $1$ the distributed policies and to $0$ the centralized one. 
Unless specified otherwise, we focus on the \textit{distributed selfish} policy.

\begin{figure}[h!t]
    \centering 
    \setlength\fwidth{0.75\columnwidth}
    \setlength\fheight{0.29\columnwidth}
    \input{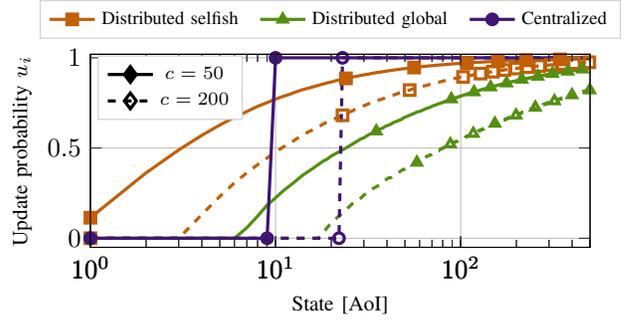}
    \caption{Update probability $u_i$ versus \gls{aoi} for the different reward functions, for $N = 10$ and $c = \{50, 200\}$.}
    \label{fig:2_sources_p_tx_vs_state}
\end{figure}

Fig.~\ref{fig:2_sources_p_tx_vs_state} shows the update probabilities $u_i$ versus the \gls{aoi} of the system $i$. We consider $N = 10$ sources and the values of the transmission cost are $c \in \{50, 200\}$. In general, a higher state, i.e., a higher \gls{aoi}, corresponds to an increased transmission probability, regardless of the specific policy considered, in accordance to the results proven in the previous section. 
Note that, given a fixed transmission probability, the expected rewards are monotonically decreasing in the \gls{aoi}. Therefore, it is convenient to transmit with a higher probability. Notice that $u_i$ is also the probability that the state is reset to $0$.

\begin{figure}
  \centering
  \begin{subfigure}[t]{\columnwidth}
    \centering 
    \setlength\fwidth{0.75\columnwidth}
    \setlength\fheight{0.29\columnwidth}
    \input{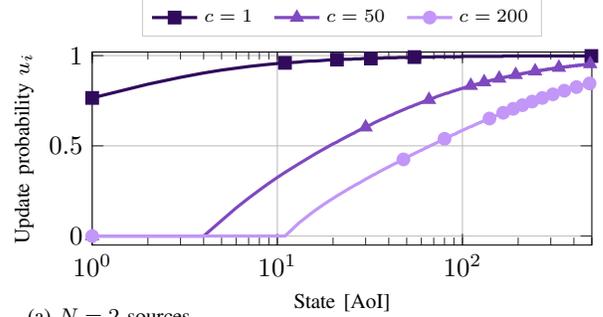}
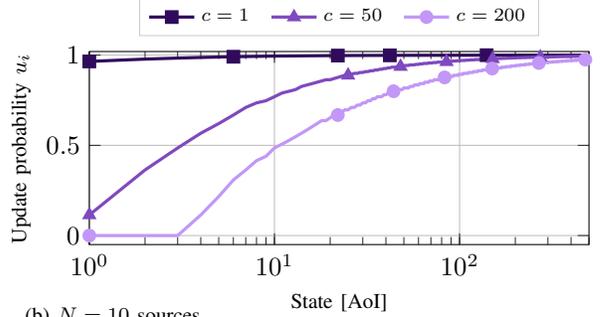

    \vspace*{-4mm}
    \caption{$N = 2$ sources. \mbox{\qquad}\mbox{\qquad}\mbox{\qquad}\mbox{\qquad}\mbox{\qquad}\mbox{\qquad}\mbox{\qquad}\mbox{\qquad}\mbox{\qquad}}
    \vspace*{2mm}
    \label{fig:2_sources_p_tx_vs_state_and_cost}
  \end{subfigure}
 \hfill
  \begin{subfigure}[t]{\columnwidth}
    \centering 
    \setlength\fwidth{0.75\columnwidth}
    \setlength\fheight{0.29\columnwidth}
    % This file was created with tikzplotlib v0.10.1.
\begin{tikzpicture}

    \definecolor{plotColor3Dark}{HTML}{300b5c}
    \definecolor{plotColor3}{HTML}{7f49bf}
    \definecolor{plotColor3Light}{HTML}{c297f7}

 \begin{semilogxaxis}[
    width=\fwidth,
    height=\fheight,
    at={(0\fwidth,0\fheight)},
    scale only axis,
    legend image post style={mark indices={}},
    legend style={
        /tikz/every even column/.append style={column sep=0.2cm},
        at={(0.5, 1.08)}, 
        anchor=south, 
        draw=white!80!black, 
        font=\scriptsize
        },
    legend columns=3,
    %tick align=outside,
    %x grid style={white!80!black},
    xlabel style={font=\footnotesize},
    xlabel={State [AoI]},
    xmajorgrids,
    %xmajorticks=false,
    xmin=1, xmax=500,
    xtick style={color=white!15!black},
    %y grid style={white!80!black},
    ylabel shift = -1 pt,
    ylabel style={font=\footnotesize},
    ylabel={Update probability $u_i$},
    ymajorgrids,
    ymajorticks=true,
    ymin=-0.05, ymax=1.02,
    ytick style={color=white!15!black}
]
\addplot [very thick, plotColor3Dark, mark=square*, mark repeat=4]
table {%
0 0.925694942474365
1 0.965149641036987
2 0.977996349334717
3 0.983675241470337
4 0.98722779750824
6 0.991263866424561
9 0.994113326072693
10 0.994857430458069
16 0.996610045433044
22 0.997637391090393
27 0.998109817504883
32 0.998375415802002
37 0.998607158660889
42 0.998808622360229
57 0.99913489818573
64 0.999265670776367
126 0.999629497528076
140 0.999690055847168
1000 0.999965906143188
};
\addlegendentry{$c = 1$}
\addplot [very thick, plotColor3, mark=triangle*, mark repeat=20]
table {%
0 0
1 0.114155292510986
2 0.363532662391663
3 0.485286235809326
4 0.567126154899597
5 0.62060558795929
6 0.668063402175903
7 0.710109949111938
8 0.735411167144775
9 0.747300863265991
10 0.769648432731628
11 0.790201425552368
12 0.799845576286316
13 0.809088945388794
14 0.826432108879089
16 0.842344164848328
18 0.856931209564209
19 0.86375880241394
20 0.86375880241394
22 0.87654173374176
24 0.888234853744507
25 0.888234853744507
27 0.898921608924866
28 0.898921608924866
29 0.90391206741333
30 0.90391206741333
31 0.908679723739624
32 0.908679723739624
33 0.913233399391174
34 0.913233399391174
35 0.917581677436829
36 0.917581677436829
37 0.921732902526855
38 0.921732902526855
39 0.925694942474365
40 0.925694942474365
41 0.929475545883179
42 0.929475545883179
43 0.933082103729248
45 0.933082103729248
46 0.936521768569946
48 0.936521768569946
49 0.939801573753357
50 0.939801573753357
51 0.942928075790405
53 0.942928075790405
54 0.945907592773438
57 0.945907592773438
58 0.9487464427948
60 0.9487464427948
61 0.95145058631897
63 0.95145058631897
64 0.954025506973267
67 0.954025506973267
68 0.956476926803589
71 0.956476926803589
72 0.958809971809387
75 0.958809971809387
76 0.961029887199402
80 0.961029887199402
81 0.963141560554504
85 0.963141560554504
86 0.965149641036987
90 0.965149641036987
91 0.967058897018433
95 0.967058897018433
96 0.968873262405396
101 0.968873262405396
102 0.970597267150879
107 0.970597267150879
108 0.972234964370728
113 0.972234964370728
114 0.973789930343628
120 0.973789930343628
121 0.975266218185425
127 0.975266218185425
128 0.976667165756226
134 0.976667165756226
135 0.977996349334717
142 0.977996349334717
143 0.979256868362427
151 0.979256868362427
152 0.980452060699463
160 0.980452060699463
161 0.981585025787354
170 0.981585025787354
171 0.982658386230469
180 0.982658386230469
181 0.983675241470337
191 0.983675241470337
192 0.984638094902039
202 0.984638094902039
203 0.985549449920654
215 0.985549449920654
216 0.986412048339844
228 0.986412048339844
229 0.98722779750824
242 0.98722779750824
243 0.987999320030212
257 0.987999320030212
258 0.988728523254395
273 0.988728523254395
274 0.989417791366577
290 0.989417791366577
291 0.990068793296814
308 0.990068793296814
309 0.990683555603027
327 0.990683555603027
328 0.991263866424561
347 0.991263866424561
348 0.991811275482178
369 0.991811275482178
370 0.992327809333801
393 0.992327809333801
394 0.992814779281616
418 0.992814779281616
419 0.993273735046387
444 0.993273735046387
445 0.993706107139587
473 0.993706107139587
474 0.994113326072693
503 0.994113326072693
504 0.994496703147888
571 0.994857430458069
573 0.99519681930542
649 0.995515823364258
651 0.995815634727478
738 0.996097326278687
740 0.99636173248291
840 0.996610045433044
842 0.996842861175537
959 0.997061252593994
961 0.997266054153442
1000 0.997266054153442
};
\addlegendentry{$c = 50$}
\addplot [very thick, plotColor3Light, mark=*, mark repeat=20]
table {%
0 0
1 0
3 0
4 0.114155292510986
5 0.216445446014404
6 0.307979583740234
7 0.363532662391663
8 0.415039539337158
9 0.439358592033386
10 0.485286235809326
11 0.506955146789551
12 0.527799606323242
13 0.547847509384155
14 0.567126154899597
15 0.585661768913269
16 0.603479981422424
17 0.62060558795929
18 0.637062549591064
19 0.637062549591064
20 0.652874350547791
21 0.668063402175903
22 0.668063402175903
23 0.68265163898468
24 0.696660280227661
25 0.696660280227661
26 0.710109949111938
27 0.710109949111938
28 0.723020553588867
29 0.723020553588867
30 0.735411167144775
31 0.735411167144775
32 0.747300863265991
33 0.747300863265991
34 0.758707404136658
35 0.758707404136658
36 0.769648432731628
38 0.769648432731628
39 0.780140995979309
40 0.780140995979309
41 0.790201425552368
43 0.790201425552368
44 0.799845576286316
46 0.799845576286316
47 0.809088945388794
49 0.809088945388794
50 0.817946314811707
53 0.817946314811707
54 0.826432108879089
56 0.826432108879089
57 0.83456015586853
60 0.83456015586853
61 0.842344164848328
64 0.842344164848328
65 0.849797010421753
68 0.849797010421753
69 0.856931209564209
72 0.856931209564209
73 0.86375880241394
77 0.86375880241394
78 0.870291948318481
82 0.870291948318481
83 0.87654173374176
87 0.87654173374176
88 0.882519125938416
92 0.882519125938416
93 0.888234853744507
98 0.888234853744507
99 0.893698930740356
104 0.893698930740356
105 0.898921608924866
111 0.898921608924866
112 0.90391206741333
118 0.90391206741333
119 0.908679723739624
125 0.908679723739624
126 0.913233399391174
132 0.913233399391174
133 0.917581677436829
141 0.917581677436829
142 0.921732902526855
149 0.921732902526855
150 0.925694942474365
158 0.925694942474365
159 0.929475545883179
168 0.929475545883179
169 0.933082103729248
178 0.933082103729248
179 0.936521768569946
189 0.936521768569946
190 0.939801573753357
200 0.939801573753357
201 0.942928075790405
212 0.942928075790405
213 0.945907592773438
225 0.945907592773438
226 0.9487464427948
238 0.9487464427948
239 0.95145058631897
252 0.95145058631897
253 0.954025506973267
267 0.954025506973267
268 0.956476926803589
283 0.956476926803589
284 0.958809971809387
300 0.958809971809387
301 0.961029887199402
318 0.961029887199402
319 0.963141560554504
337 0.963141560554504
338 0.965149641036987
357 0.965149641036987
358 0.967058897018433
378 0.967058897018433
379 0.968873262405396
401 0.968873262405396
402 0.970597267150879
425 0.970597267150879
426 0.972234964370728
450 0.972234964370728
451 0.973789930343628
477 0.973789930343628
478 0.975266218185425
505 0.975266218185425
506 0.976667165756226
536 0.976667165756226
537 0.977996349334717
568 0.977996349334717
569 0.979256868362427
602 0.979256868362427
603 0.980452060699463
638 0.980452060699463
639 0.981585025787354
677 0.981585025787354
678 0.982658386230469
718 0.982658386230469
719 0.983675241470337
761 0.983675241470337
762 0.984638094902039
808 0.984638094902039
809 0.985549449920654
857 0.985549449920654
858 0.986412048339844
910 0.986412048339844
911 0.98722779750824
966 0.98722779750824
967 0.987999320030212
1000 0.987999320030212
};
\addlegendentry{$c = 200$}
\end{semilogxaxis}

\end{tikzpicture}
    \vspace*{-4mm}
    \caption{$N = 10$ sources.\mbox{\qquad}\mbox{\qquad}\mbox{\qquad}\mbox{\qquad}\mbox{\qquad}\mbox{\qquad}\mbox{\qquad}\mbox{\qquad}\mbox{\qquad}}
    \label{fig:10_sources_p_tx_vs_state_and_cost}
  \end{subfigure}
  \caption{Update probability $u_i$ versus \gls{aoi} for different values of the transmission cost $c$, $N = \{2, 10\}$ sources, \textit{distributed selfish} system.}
  \vspace{-0.3cm}
  \label{Fig:p_tx_vs_state_and_cost}
\end{figure}

If the cost is increased, all curves shift to the right, which corresponds to a lower update probability. However, while this and all qualitative behaviors of the three policies follow the same monotonic trend, the shapes are different. 
This agrees with the theoretical results found, since the \textit{centralized} policy has a sharp transition once a threshold value is reached. Distributed approaches have instead a smooth increase around the same threshold. It can also be noticed that the \textit{distributed selfish} policy leads to higher update probabilities compared to the \textit{distributed global} policy, since the cost incurred by other sources is neglected, thus a more aggressive update is obtained. This is not necessarily more efficient, since a higher transmission cost is paid (and notably, the nodes are not coordinated so redundant transmissions may happen).

Figs.~\ref{fig:2_sources_p_tx_vs_state_and_cost} and~\ref{fig:10_sources_p_tx_vs_state_and_cost} depict the update probabilities as a function of the transmission cost $c$ 
%for a fixed reward function, i.e., 
within a \textit{distributed selfish} system. For sufficiently low transmission costs, the optimal update probability is non-zero even in the initial state. On the other hand, as the cost increases $\frac{\partial \bar{K}(i,\pi)}{p_i}$ becomes negative for all states below an age threshold. Thus, the optimal strategy is to never transmit an update until such a state is visited, thus corroborating the theoretical results of Sec.~\ref{sec:theo}. Finally, it can also be seen how the update probability is directly proportional to the number of sources, which is to be expected as the \textit{distributed selfish} reward function exhibits a myopic cost term which takes into account the individual transmission cost only.

%\begin{figure}[h!t]
%    \centering 
%    \setlength\fwidth{0.8\columnwidth}
%    \setlength\fheight{0.37\columnwidth}
%    \input{figures/2_sources_p_tx_vs_state_and_cost}
%    \caption{Update probability versus \gls{aoi} for different values of the cost $c$, considering $N = 2$ sources and the \textit{distributed selfish} system.}
%    \label{fig:2_sources_p_tx_vs_state_and_cost}
%\end{figure}

%\begin{figure}[h!t]
%    \centering 
%    \setlength\fwidth{0.8\columnwidth}
%    \setlength\fheight{0.37\columnwidth}
%    \input{figures/10_sources_p_tx_vs_state_and_cost}
%    \caption{Transmission probability versus \gls{aoi} for different values of the cost $c$, considering $N = 10$ sources and the distributed selfish system.}
%    \label{fig:10_sources_p_tx_vs_state_and_cost}
%\end{figure}

%\begin{figure}[h!t]
%    \centering 
%    \setlength\fwidth{0.8\columnwidth}
%    \setlength\fheight{0.37\columnwidth}
%    \input{figures/2_sources_avg_p_tx_cost}
%    \caption{Average transmission probability versus cost $c$ for different reward functions, considering $N = 2$ sources.}
%    \label{fig:2_sources_avg_p_tx__cost}
%\end{figure}

%\begin{figure}[h!t]
%    \centering 
%    \setlength\fwidth{0.8\columnwidth}
%    \setlength\fheight{0.4\columnwidth}
%    \input{figures/10_sources_avg_p_tx_cost}
%    \caption{Average transmission probability versus cost $c$ for different reward functions, considering $N = 10$ sources.}
%    \label{fig:10_sources_avg_p_tx__cost}
%\end{figure}

\begin{figure}
  \centering
  \begin{subfigure}[t]{\columnwidth}
    \centering 
    \setlength\fwidth{0.75\columnwidth}
    \setlength\fheight{0.29\columnwidth}
    % This file was created with tikzplotlib v0.10.1.
\begin{tikzpicture}

  \definecolor{plotColor1}{HTML}{C25F08}
  \definecolor{plotColor2}{HTML}{548F10}
  \definecolor{plotColor3}{HTML}{411475}

 \begin{axis}[
    width=\fwidth,
    height=\fheight,
    at={(0\fwidth,0\fheight)},
    scale only axis,
    legend image post style={mark indices={}},
    legend style={
        /tikz/every even column/.append style={column sep=0.2cm},
        at={(0.5, 1.08)}, 
        anchor=south, 
        draw=white!80!black, 
        font=\scriptsize
        },
    legend columns=3,
    %tick align=outside,
    %x grid style={white!80!black},
    xlabel style={font=\footnotesize},
    xlabel={Transmission cost $c$},
    xmajorgrids,
    %xmajorticks=false,
    xmin=10, xmax=190,
    xtick style={color=white!15!black},
    %y grid style={white!80!black},
    ylabel shift = -1 pt,
    ylabel style={font=\footnotesize},
    ylabel={Average update probability $\bar{U}$},
    ymajorgrids,
    ymajorticks=true,
    ymin=0.0, ymax=0.3,
    ytick style={color=white!15!black}
]
\addplot [very thick, plotColor1, mark=square*, mark repeat=12]
table {%
10 0.266861438751221
12 0.245872378349304
18 0.204259157180786
20 0.191392779350281
22 0.183743596076965
26 0.171010136604309
28 0.165204286575317
30 0.159098029136658
32 0.153120636940002
34 0.147848129272461
36 0.14432954788208
38 0.141159057617188
40 0.137755393981934
42 0.134641885757446
44 0.131128907203674
46 0.12777042388916
48 0.124281048774719
52 0.119963645935059
54 0.118160963058472
58 0.114252209663391
60 0.111899733543396
62 0.109849691390991
64 0.107508540153503
66 0.105966448783875
68 0.104685068130493
70 0.103127956390381
78 0.0976927280426025
80 0.0962469577789307
82 0.0950647592544556
84 0.0934542417526245
88 0.091494083404541
90 0.0906832218170166
92 0.089781641960144
94 0.0885647535324097
96 0.0878877639770508
98 0.0866469144821167
100 0.0855803489685059
102 0.0846767425537109
104 0.0835722684860229
106 0.0828410387039185
108 0.0822861194610596
110 0.0816217660903931
112 0.0806480646133423
114 0.0800026655197144
116 0.0792000293731689
118 0.078569769859314
124 0.0761486291885376
126 0.075481653213501
130 0.0745220184326172
132 0.073830246925354
144 0.0704944133758545
146 0.0697165727615356
148 0.0692138671875
152 0.0684490203857422
160 0.0665915012359619
170 0.0643895864486694
172 0.0638115406036377
176 0.0631272792816162
180 0.06252121925354
186 0.0613067150115967
190 0.0607326030731201
196 0.059529185295105
198 0.0591801404953003
};
\addlegendentry{Distributed selfish}
\addplot [very thick, plotColor2, mark=triangle*, mark repeat=12]
table {%
10 0.191392779350281
12 0.177412986755371
14 0.165204286575317
16 0.153120636940002
18 0.14432954788208
22 0.131128907203674
24 0.124281048774719
26 0.119963645935059
28 0.116178274154663
30 0.111899733543396
32 0.107508540153503
36 0.101786375045776
38 0.099135160446167
40 0.0962469577789307
42 0.0934542417526245
44 0.091494083404541
46 0.089781641960144
48 0.0878877639770508
50 0.0855803489685059
52 0.0835722684860229
54 0.0822861194610596
56 0.0806480646133423
60 0.0777758359909058
62 0.0761486291885376
66 0.073830246925354
72 0.0704944133758545
74 0.0692138671875
76 0.0684490203857422
82 0.0656696557998657
84 0.0648120641708374
86 0.0638115406036377
96 0.0603189468383789
100 0.058762788772583
102 0.0583269596099854
104 0.0577600002288818
106 0.0570653676986694
108 0.056460976600647
110 0.0559635162353516
112 0.0552250146865845
122 0.0528402328491211
124 0.0521782636642456
126 0.051615834236145
134 0.0500771999359131
136 0.0496424436569214
138 0.0490951538085938
142 0.0483359098434448
144 0.048051118850708
148 0.0473396778106689
152 0.0466599464416504
158 0.0454955101013184
162 0.0449281930923462
164 0.0445399284362793
166 0.044312596321106
170 0.0436273813247681
182 0.0419967174530029
184 0.04163658618927
188 0.0411460399627686
194 0.0404614210128784
196 0.0402270555496216
198 0.0398954153060913
};
\addlegendentry{Distributed global}
\addplot [very thick, plotColor3, mark=*, mark repeat=20]
table {%
10 0.200000047683716
14 0.200000047683716
16 0.166666626930237
18 0.166666626930237
20 0.142857193946838
24 0.142857193946838
26 0.125
32 0.125
34 0.111111164093018
38 0.111111164093018
40 0.100000023841858
46 0.100000023841858
48 0.0909091234207153
56 0.0909091234207153
58 0.0833333730697632
64 0.0833333730697632
66 0.076923131942749
74 0.076923131942749
76 0.0714285373687744
84 0.0714285373687744
86 0.0666667222976685
96 0.0666667222976685
98 0.0625
106 0.0625
108 0.0588235855102539
118 0.0588235855102539
120 0.0555555820465088
130 0.0555555820465088
132 0.0526316165924072
142 0.0526316165924072
144 0.0499999523162842
156 0.0499999523162842
158 0.047619104385376
168 0.047619104385376
170 0.0454545021057129
182 0.0454545021057129
184 0.04347825050354
196 0.04347825050354
198 0.0416666269302368
};
\addlegendentry{Centralized}
\end{axis}

\end{tikzpicture}
    \vspace*{-4mm}
    \caption{$N = 2$ sources.\mbox{\qquad}\mbox{\qquad}\mbox{\qquad}\mbox{\qquad}\mbox{\qquad}\mbox{\qquad}\mbox{\qquad}\mbox{\qquad}\mbox{\qquad}}
    \vspace*{2mm}
    \label{fig:2_sources_avg_p_tx_cost}
  \end{subfigure}
 \hfill
  \begin{subfigure}[t]{\columnwidth}
    \centering 
    \setlength\fwidth{0.75\columnwidth}
    \setlength\fheight{0.29\columnwidth}
    % This file was created with tikzplotlib v0.10.1.
\begin{tikzpicture}

  \definecolor{plotColor1}{HTML}{C25F08}
  \definecolor{plotColor2}{HTML}{548F10}
  \definecolor{plotColor3}{HTML}{411475}

 \begin{axis}[
    width=\fwidth,
    height=\fheight,
    at={(0\fwidth,0\fheight)},
    scale only axis,
    legend image post style={mark indices={}},
    legend style={
        /tikz/every even column/.append style={column sep=0.2cm},
        at={(0.5, 1.08)}, 
        anchor=south, 
        draw=white!80!black, 
        font=\scriptsize
        },
    legend columns=3,
    %tick align=outside,
    %x grid style={white!80!black},
    xlabel style={font=\footnotesize},
    xlabel={Transmission cost $c$},
    xmajorgrids,
    %xmajorticks=false,
    xmin=10, xmax=190,
    xtick style={color=white!15!black},
    %y grid style={white!80!black},
    ylabel shift = -1 pt,
    ylabel style={font=\footnotesize},
    ylabel={Average update probability $\bar{U}$},
    ymajorgrids,
    ymajorticks=true,
    ymin=0.0, ymax=0.6,
    ytick style={color=white!15!black}
]

\addplot [very thick, plotColor1, mark=square*, mark repeat=12]
table {%
10 0.540003299713135
12 0.485310316085815
14 0.450286507606506
16 0.419069528579712
18 0.38816511631012
20 0.370621562004089
22 0.34996771812439
24 0.341284871101379
26 0.33403491973877
28 0.326541304588318
30 0.318760395050049
32 0.31069278717041
34 0.302334189414978
36 0.293689489364624
38 0.292260766029358
40 0.283290863037109
42 0.275510549545288
44 0.273765206336975
46 0.264259696006775
48 0.26129412651062
50 0.254119753837585
52 0.246155738830566
54 0.243844151496887
56 0.240892171859741
58 0.23333203792572
60 0.233331322669983
62 0.22287917137146
64 0.222578048706055
66 0.219374656677246
68 0.216515302658081
70 0.216513991355896
72 0.212302684783936
74 0.210127592086792
76 0.209957957267761
78 0.205297827720642
80 0.203409552574158
82 0.203207731246948
84 0.198827147483826
86 0.196352362632751
88 0.196339964866638
90 0.195981860160828
92 0.191019177436829
94 0.188934922218323
98 0.188501358032227
100 0.183357238769531
102 0.181184053421021
104 0.180818319320679
106 0.180668115615845
108 0.175346732139587
110 0.173090696334839
112 0.173089146614075
114 0.17247998714447
116 0.169128775596619
120 0.164661884307861
122 0.164164304733276
124 0.163945317268372
126 0.163945317268372
130 0.159244775772095
132 0.159139633178711
134 0.15840744972229
138 0.158398985862732
142 0.153472542762756
144 0.153343915939331
146 0.152498126029968
152 0.15241265296936
156 0.147332549095154
158 0.147141695022583
160 0.146203637123108
166 0.146091103553772
168 0.143293023109436
170 0.140814781188965
172 0.140580058097839
174 0.139514684677124
182 0.139369010925293
184 0.133915185928345
188 0.133628726005554
190 0.132847905158997
192 0.132397651672363
198 0.132240533828735
};
\addlegendentry{Distributed selfish}
\addplot [very thick, plotColor2, mark=triangle*, mark repeat=12]
table {%
10 0.183357238769531
12 0.164661884307861
14 0.155876874923706
16 0.146203637123108
18 0.139369010925293
20 0.12913990020752
22 0.123985648155212
24 0.119845151901245
26 0.115912318229675
28 0.109898686408997
30 0.10726523399353
32 0.103268265724182
34 0.100063800811768
36 0.097739577293396
38 0.0944652557373047
40 0.0936340093612671
42 0.0900603532791138
44 0.0890858173370361
46 0.0851949453353882
48 0.0845509767532349
50 0.0824147462844849
54 0.0799456834793091
56 0.0768642425537109
60 0.0753439664840698
62 0.0731747150421143
64 0.0723140239715576
66 0.0722168684005737
68 0.0702011585235596
70 0.0689312219619751
76 0.0661959648132324
78 0.0644506216049194
80 0.0643634796142578
82 0.0633957386016846
84 0.0631271600723267
86 0.0612442493438721
90 0.0610493421554565
94 0.0589827299118042
96 0.0575834512710571
98 0.0573444366455078
100 0.0573420524597168
102 0.0570318698883057
106 0.0547604560852051
108 0.0541379451751709
110 0.0538996458053589
112 0.053895115852356
116 0.0530391931533813
118 0.0513182878494263
128 0.0501763820648193
130 0.0496175289154053
132 0.0484638214111328
134 0.0482182502746582
136 0.0482178926467896
138 0.0476493835449219
140 0.0476493835449219
142 0.0472481250762939
146 0.0457606315612793
148 0.0457465648651123
150 0.0454633235931396
152 0.0454342365264893
154 0.0448242425918579
156 0.0447068214416504
164 0.043124794960022
166 0.043102502822876
168 0.042759895324707
170 0.0427592992782593
174 0.0413120985031128
178 0.0413117408752441
180 0.0409619808197021
182 0.0409398078918457
184 0.0406482219696045
186 0.0401235818862915
188 0.0400815010070801
190 0.0396133661270142
192 0.0396120548248291
196 0.0389258861541748
198 0.0388778448104858
};
\addlegendentry{Distributed global}
\addplot [very thick, plotColor3, mark=*, mark repeat=12]
table {%
10 0.200000047683716
14 0.200000047683716
16 0.166666626930237
18 0.166666626930237
20 0.142857193946838
24 0.142857193946838
26 0.125
32 0.125
34 0.111111164093018
38 0.111111164093018
40 0.100000023841858
46 0.100000023841858
48 0.0909091234207153
56 0.0909091234207153
58 0.0833333730697632
64 0.0833333730697632
66 0.076923131942749
74 0.076923131942749
76 0.0714285373687744
84 0.0714285373687744
86 0.0666667222976685
96 0.0666667222976685
98 0.0625
106 0.0625
108 0.0588235855102539
118 0.0588235855102539
120 0.0555555820465088
130 0.0555555820465088
132 0.0526316165924072
142 0.0526316165924072
144 0.0499999523162842
156 0.0499999523162842
158 0.047619104385376
168 0.047619104385376
170 0.0454545021057129
182 0.0454545021057129
184 0.04347825050354
196 0.04347825050354
198 0.0416666269302368
};
\addlegendentry{Centralized}
\end{axis}

\end{tikzpicture}
    \vspace*{-4mm}
    \caption{$N = 10$ sources.\mbox{\qquad}\mbox{\qquad}\mbox{\qquad}\mbox{\qquad}\mbox{\qquad}\mbox{\qquad}\mbox{\qquad}\mbox{\qquad}\mbox{\qquad}}
    \label{fig:10_sources_avg_p_tx_cost}
  \end{subfigure}
    \caption{Average update probability $\bar{U}$ versus cost $c$ for different reward functions, considering $N = \{2, 10\}$ sources.}
  \label{Fig:avg_p_tx_vs_cost}
\end{figure}
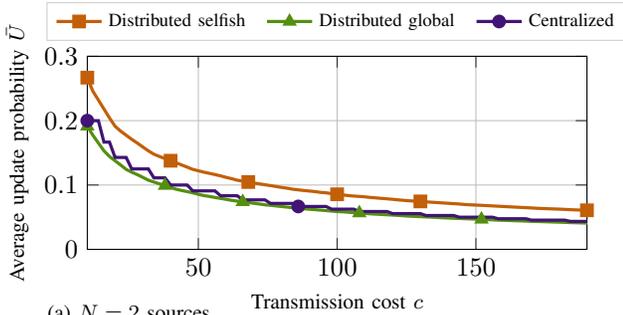
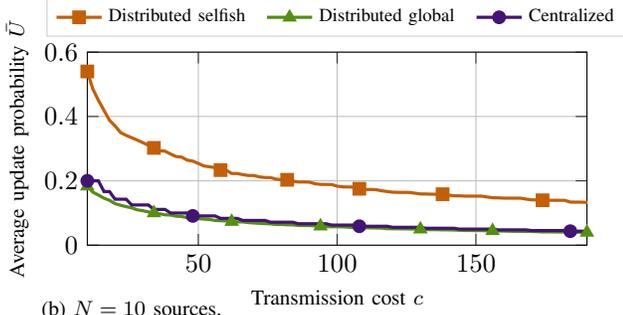

Figs.~\ref{fig:2_sources_avg_p_tx_cost} and~\ref{fig:10_sources_avg_p_tx_cost} report the average update probability of the system $\bar{U}$ for $N = 2$ and $N = 10$ sources, respectively. Throughout the whole range of considered costs $c$, the \textit{distributed global} and \textit{centralized} policies lead to similar update probabilities, with the latter transmitting slightly more often. Conversely, the \textit{distributed selfish} policy results in a more aggressive source behavior. The gap in terms of update probability with respect to the competitors is proportional to $N$, validating our choice of comparing the different policies only for low values of $N$. Furthermore, the \textit{distributed global} (\textit{centralized}) policy leads to solutions which show little (no) variability with respect to $N$, which suggests that these approaches are robust to estimation errors of $N$.

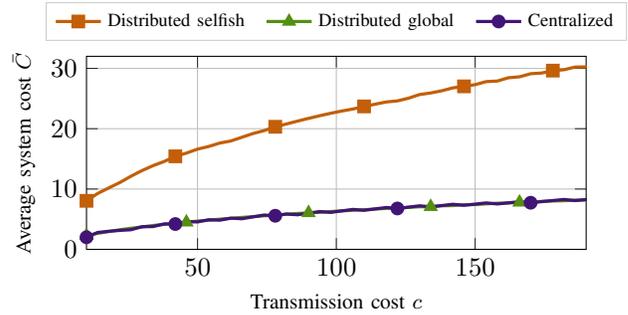
\begin{figure}[t]
    \centering 
    \setlength\fwidth{0.75\columnwidth}
    \setlength\fheight{0.29\columnwidth}
    % This file was created with tikzplotlib v0.10.1.
\begin{tikzpicture}

  \definecolor{plotColor1}{HTML}{C25F08}
  \definecolor{plotColor2}{HTML}{548F10}
  \definecolor{plotColor3}{HTML}{411475}

 \begin{axis}[
    width=\fwidth,
    height=\fheight,
    at={(0\fwidth,0\fheight)},
    scale only axis,
    legend image post style={mark indices={}},
    legend style={
        /tikz/every even column/.append style={column sep=0.2cm},
        at={(0.5, 1.08)}, 
        anchor=south, 
        draw=white!80!black, 
        font=\scriptsize
        },
    legend columns=3,
    %tick align=outside,
    %x grid style={white!80!black},
    xlabel style={font=\footnotesize},
    xlabel={Transmission cost $c$},
    xmajorgrids,
    %xmajorticks=false,
    xmin=10, xmax=190,
    xtick style={color=white!15!black},
    %y grid style={white!80!black},
    ylabel shift = -1 pt,
    ylabel style={font=\footnotesize},
    ylabel={Average system cost $\bar{C}$},
    ymajorgrids,
    ymajorticks=true,
    ymin=0, ymax=32,
    ytick style={color=white!15!black}
]

\addplot [very thick, plotColor1, mark=square*, mark repeat=8]
table {%
10 8.0473575592041
14 9.25468826293945
18 10.1784620285034
22 11.0717582702637
26 12.1186237335205
30 13.0585927963257
34 13.8801307678223
38 14.5712413787842
42 15.422532081604
46 15.9564161300659
50 16.6163463592529
54 17.0509033203125
58 17.6227283477783
62 17.9657382965088
70 19.2042770385742
74 19.6891288757324
78 20.3228282928467
82 20.7807197570801
86 21.255521774292
90 21.7131423950195
94 22.1465263366699
98 22.5585918426514
102 22.9343357086182
106 23.2661190032959
110 23.6936702728271
118 24.4366931915283
122 24.6079845428467
126 25.0563526153564
130 25.6660137176514
134 25.9240894317627
138 26.2889385223389
142 26.7719402313232
146 27.021842956543
150 27.3021392822266
154 27.8097019195557
158 27.9009628295898
162 28.4469528198242
166 28.6007556915283
170 29.1126937866211
174 29.2262096405029
178 29.6306095123291
182 29.7731304168701
186 30.2177066802979
190 30.2470035552979
194 30.7758045196533
198 30.8357181549072
};
\addlegendentry{Distributed selfish}
\addplot [very thick, plotColor2, mark=triangle*, mark repeat=8]
table {%
10 2.28546953201294
14 2.64991497993469
18 2.94496822357178
22 3.25636792182922
30 3.72346997261047
34 3.961505651474
38 4.14069366455078
42 4.3385272026062
46 4.49461650848389
50 4.69532203674316
54 4.82436180114746
58 5.00980663299561
70 5.43280076980591
78 5.68610239028931
82 5.83007526397705
86 5.94737863540649
90 6.05185651779175
94 6.13252401351929
98 6.26773738861084
102 6.39044094085693
106 6.46524620056152
118 6.78533363342285
122 6.83609914779663
130 7.0627384185791
134 7.06437015533447
138 7.22194671630859
142 7.25672626495361
146 7.40202713012695
150 7.42247581481934
154 7.54171419143677
158 7.64461278915405
162 7.6834397315979
166 7.79500722885132
170 7.81441020965576
174 7.95347738265991
178 8.02476119995117
182 8.05392169952393
186 8.14624118804932
190 8.16541290283203
198 8.35414505004883
};
\addlegendentry{Distributed global}
\addplot [very thick, plotColor3, mark=*, mark repeat=8]
table {%
10 2
14 2.79999995231628
18 3
22 3.14285707473755
26 3.25
30 3.75
34 3.77777767181396
38 4.22222232818604
42 4.19999980926514
46 4.59999990463257
50 4.54545450210571
54 4.90909099578857
58 4.83333349227905
62 5.16666650772095
66 5.07692289352417
74 5.692307472229
78 5.57142877578735
82 5.85714292526245
86 5.73333311080933
94 6.26666688919067
98 6.125
106 6.625
110 6.47058820724487
118 6.94117641448975
122 6.77777767181396
130 7.22222232818604
134 7.05263137817383
142 7.47368431091309
146 7.30000019073486
154 7.69999980926514
158 7.5238094329834
166 7.90476179122925
170 7.72727251052856
182 8.27272701263428
186 8.08695697784424
194 8.43478298187256
198 8.25
};
\addlegendentry{Centralized}
\end{axis}

\end{tikzpicture}
\vspace*{-0.2cm}
    \caption{Expected system cost $\bar{C}$ versus transmission cost $c$ for different reward functions, considering $N = 10$ sources. }
    \vspace{-0.3cm}
  \label{fig:10_sources_avg_cost_vs_cost}
\end{figure}

\begin{figure}[t]
    \centering 
    \setlength\fwidth{0.75\columnwidth}
    \setlength\fheight{0.29\columnwidth}
    % This file was created with tikzplotlib v0.10.1.
\begin{tikzpicture}

  \definecolor{plotColor1}{HTML}{C25F08}
  \definecolor{plotColor2}{HTML}{548F10}
  \definecolor{plotColor3}{HTML}{411475}

 \begin{axis}[
    width=\fwidth,
    height=\fheight,
    at={(0\fwidth,0\fheight)},
    scale only axis,
    legend image post style={mark indices={}},
    legend style={
        /tikz/every even column/.append style={column sep=0.2cm},
        at={(0.5, 1.08)}, 
        anchor=south, 
        draw=white!80!black, 
        font=\scriptsize
        },
    legend columns=3,
    %tick align=outside,
    %x grid style={white!80!black},
    xlabel style={font=\footnotesize},
    xlabel={Transmission cost $c$},
    xmajorgrids,
    %xmajorticks=false,
    xmin=10, xmax=190,
    xtick style={color=white!15!black},
    %y grid style={white!80!black},
    ylabel shift = -1 pt,
    ylabel style={font=\footnotesize},
    ylabel={Average \gls{aoi} $\bar{\Delta}$},
    ymajorgrids,
    ymajorticks=true,
    ymin=0, ymax=14,
    ytick style={color=white!15!black}
]

\addplot [very thick, plotColor1, mark=square*, mark repeat=12]
table {%
10 0.62944495677948
12 0.720275163650513
14 0.804237723350525
16 0.879419088363647
18 0.96070921421051
20 1.03194558620453
22 1.08819365501404
28 1.22521817684174
34 1.36081576347351
36 1.39826023578644
38 1.45488893985748
40 1.4851781129837
42 1.52333188056946
44 1.58158016204834
50 1.69129586219788
52 1.72981202602386
54 1.78265595436096
56 1.8159567117691
58 1.85390746593475
62 1.94430410861969
64 1.96413064002991
66 1.99983537197113
68 2.02271914482117
70 2.05290198326111
72 2.07808351516724
74 2.11676430702209
76 2.13482880592346
78 2.16179656982422
80 2.20392274856567
82 2.22323799133301
86 2.28000831604004
88 2.30428004264832
90 2.33612084388733
92 2.35575008392334
94 2.39248728752136
96 2.40993523597717
98 2.4489426612854
100 2.46651387214661
102 2.50720810890198
104 2.52561092376709
106 2.56821799278259
108 2.58749866485596
110 2.61704301834106
112 2.65231418609619
114 2.66956734657288
116 2.70230937004089
118 2.72025108337402
120 2.76925587654114
122 2.79136800765991
124 2.7914035320282
126 2.83012413978577
128 2.84796738624573
130 2.85029435157776
132 2.88941287994385
134 2.9076406955719
136 2.91835522651672
138 2.95188784599304
140 2.97084784507751
142 2.98230910301208
144 3.01788425445557
146 3.0372953414917
148 3.04121327400208
150 3.08754920959473
152 3.10743093490601
154 3.11187267303467
156 3.14327454566956
158 3.18147897720337
160 3.18523478507996
162 3.1989324092865
164 3.23876619338989
166 3.25967836380005
168 3.26537990570068
170 3.27893733978271
172 3.32150650024414
174 3.34228277206421
176 3.35170578956604
178 3.37147378921509
180 3.42952537536621
182 3.42992901802063
186 3.45285034179688
188 3.52170634269714
190 3.52223777770996
192 3.53385829925537
194 3.53386616706848
196 3.59716033935547
198 3.59778428077698
};
\addlegendentry{Distributed selfish}
\addplot [very thick, plotColor2, mark=triangle*, mark repeat=12]
table {%
10 2.46651387214661
12 2.76925587654114
14 2.97084784507751
16 3.18523478507996
18 3.42952537536621
22 3.77987313270569
24 3.95578932762146
26 4.16517066955566
28 4.31321620941162
30 4.48705291748047
32 4.60949754714966
34 4.77058076858521
36 4.9429178237915
38 5.09073066711426
40 5.22047901153564
42 5.36072969436646
44 5.51242923736572
46 5.65852832794189
48 5.72981691360474
50 5.87833547592163
52 6.0122275352478
54 6.16978454589844
56 6.24824666976929
58 6.3738055229187
60 6.51060724258423
62 6.61593437194824
64 6.72700500488281
66 6.84315729141235
68 6.96871709823608
70 7.06873512268066
72 7.16030550003052
74 7.29295778274536
76 7.39381742477417
78 7.50923013687134
80 7.66155242919922
82 7.69260263442993
84 7.86670160293579
86 7.90020704269409
88 8.0510778427124
92 8.18640518188477
94 8.35912799835205
102 8.69184684753418
104 8.82652378082275
106 8.92097091674805
108 8.96976947784424
110 9.10528564453125
112 9.17723178863525
116 9.37915802001953
118 9.44209003448486
120 9.57412910461426
122 9.6838207244873
124 9.73693180084229
126 9.82649898529053
128 9.94855403900146
130 9.97603511810303
132 10.0407238006592
134 10.271312713623
136 10.2738790512085
138 10.3443098068237
140 10.3795366287231
142 10.5831851959229
144 10.5859155654907
146 10.665337562561
148 10.8004264831543
150 10.9191570281982
152 10.9397077560425
156 11.1218395233154
158 11.1592435836792
160 11.2803564071655
162 11.3757200241089
164 11.4384326934814
166 11.4873332977295
168 11.5790596008301
170 11.7257127761841
172 11.7775650024414
174 11.7885761260986
176 11.8964996337891
178 11.9483127593994
180 12.1162271499634
182 12.1655740737915
184 12.1786651611328
186 12.289852142334
188 12.3519248962402
190 12.5157928466797
192 12.5559215545654
194 12.6260108947754
198 12.7417984008789
};
\addlegendentry{Distributed global}
\addplot [very thick, plotColor3, mark=*, mark repeat=12]
table {%
10 2
14 2
16 2.5
18 2.5
20 3
24 3
26 3.5
32 3.5
34 4
38 4
40 4.5
46 4.5
48 5
56 5
58 5.5
64 5.5
66 6
74 6
76 6.5
84 6.5
86 7
96 7
98 7.5
106 7.5
108 8
118 8
120 8.5
130 8.5
132 9
142 9
144 9.5
156 9.5
158 10
168 10
170 10.5
182 10.5
184 11
196 11
198 11.5
};
\addlegendentry{Centralized}
\end{axis}

\end{tikzpicture}
    \caption{Average \gls{aoi} $\bar{\Delta}$ versus cost $c$ for different reward functions, considering $N = 10$ sources. }
    \label{fig:10_sources_avg_aoi_vs_cost}
\vspace*{-0.3cm}
\end{figure}
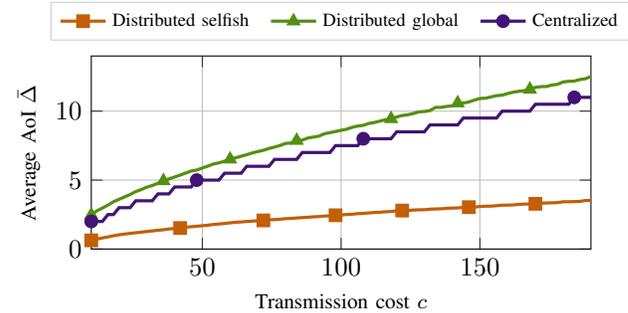

Fig.~\ref{fig:10_sources_avg_cost_vs_cost} presents the average cost $\bar{C}$ incurred by the whole system. It can be noticed that, as expected, the \textit{distributed selfish} policy leads to the highest cost. Indeed, in this case the sources consider their individual cost only, thus overestimating the optimal transmission probability. Conversely, the \textit{distributed global} and \textit{centralized} policies result in similarly lower costs, despite the different reward functions. In fact, even though in (\ref{eq:reward_global}) the cost term is proportional to $N$, the probability of transitioning to a higher \gls{aoi} decreases exponentially with respect to the number of sources as well. 
The interplay between these two phenomena leads to an update probability which is similar to the \textit{centralized} case.

The myopic behavior of the \textit{distributed selfish} policy can be seen also in Fig.~\ref{fig:10_sources_avg_aoi_vs_cost}, which reports the average \gls{aoi} of the system 
$\bar{\Delta}$. In fact, the \textit{distributed selfish} policy achieves an \gls{aoi} which is approximately 20$\%$ lower than the competitors. However, this decrease in \gls{aoi} is obtained inefficiently and thus leads to the worst system performance.

\begin{figure}[t]
    \centering 
    \setlength\fwidth{0.75\columnwidth}
    \setlength\fheight{0.29\columnwidth}
    % This file was created with tikzplotlib v0.10.1.
\begin{tikzpicture}

  \definecolor{plotColor1}{HTML}{C25F08}
  \definecolor{plotColor2}{HTML}{548F10}
  \definecolor{plotColor3}{HTML}{411475}

 \begin{axis}[
    width=\fwidth,
    height=\fheight,
    at={(0\fwidth,0\fheight)},
    scale only axis,
    legend image post style={mark indices={}},
    legend style={
        /tikz/every even column/.append style={column sep=0.2cm},
        at={(0.5, 1.08)}, 
        anchor=south, 
        draw=white!80!black, 
        font=\scriptsize
        },
    legend columns=3,
    %tick align=outside,
    %x grid style={white!80!black},
    xlabel style={font=\footnotesize},
    xlabel={Transmission cost $c$},
    xmajorgrids,
    %xmajorticks=false,
    xmin=10, xmax=190,
    xtick style={color=white!15!black},
    %y grid style={white!80!black},
    ylabel shift = -1 pt,
    ylabel style={font=\footnotesize},
    ylabel={Average system reward $\bar{\mathcal{R}}$},
    ymajorgrids,
    ymajorticks=true,
    ymin=-37.0, ymax=0.0,
    ytick style={color=white!15!black}
]

\addplot [very thick, plotColor1, mark=square*, mark repeat=9]
table {%
10 -8.67680263519287
14 -10.0589256286621
18 -11.1391706466675
22 -12.159951210022
26 -13.2991771697998
30 -14.3288326263428
34 -15.24094581604
38 -16.0261306762695
42 -16.9458637237549
46 -17.5738697052002
50 -18.3076419830322
54 -18.8335609436035
58 -19.476634979248
62 -19.910041809082
70 -21.2571792602539
74 -21.8058929443359
78 -22.4846248626709
82 -23.0039577484131
86 -23.535530090332
90 -24.0492630004883
94 -24.5390148162842
98 -25.0075340270996
102 -25.4415454864502
106 -25.8343372344971
110 -26.3107128143311
118 -27.1569442749023
122 -27.3993530273438
126 -27.8864765167236
130 -28.5163078308105
134 -28.831729888916
138 -29.2408256530762
142 -29.7542495727539
146 -30.0591373443604
150 -30.3896884918213
154 -30.9215755462646
158 -31.0824413299561
162 -31.6458854675293
166 -31.8604354858398
170 -32.3916320800781
174 -32.5684928894043
178 -33.002082824707
182 -33.2030601501465
186 -33.6705589294434
190 -33.7692413330078
194 -34.3096733093262
198 -34.4335021972656
};
\addlegendentry{Distributed selfish}
\addplot [very thick, plotColor2, mark=triangle*, mark repeat=9]
table {%
10 -4.75198316574097
14 -5.62076282501221
18 -6.37449359893799
22 -7.03624105453491
26 -7.65002679824829
30 -8.21052265167236
34 -8.73208618164062
38 -9.23142433166504
42 -9.69925689697266
46 -10.1531448364258
54 -10.9941463470459
62 -11.7686376571655
70 -12.5015354156494
78 -13.1953325271606
90 -14.1692152023315
94 -14.4916515350342
110 -15.6749429702759
122 -16.5199203491211
130 -17.0387744903564
134 -17.3356819152832
138 -17.5662574768066
142 -17.8399105072021
146 -18.0673656463623
150 -18.3416328430176
158 -18.8038558959961
162 -19.0591583251953
166 -19.282341003418
170 -19.5401229858398
174 -19.7420539855957
178 -19.9730739593506
182 -20.2194957733154
186 -20.436092376709
190 -20.6812057495117
198 -21.0959434509277
};
\addlegendentry{Distributed global}
\addplot [very thick, plotColor3, mark=*, mark repeat=9]
table {%
10 -4
14 -4.80000019073486
18 -5.5
22 -6.14285707473755
26 -6.75
30 -7.25
34 -7.77777767181396
38 -8.22222232818604
42 -8.69999980926514
46 -9.10000038146973
50 -9.54545497894287
54 -9.90909099578857
58 -10.3333330154419
62 -10.6666669845581
66 -11.0769233703613
74 -11.692307472229
78 -12.0714282989502
82 -12.3571424484253
86 -12.7333335876465
94 -13.2666664123535
98 -13.625
106 -14.125
110 -14.470588684082
118 -14.9411764144897
122 -15.277777671814
130 -15.722222328186
134 -16.0526313781738
142 -16.4736843109131
146 -16.7999992370605
154 -17.2000007629395
158 -17.5238094329834
166 -17.9047622680664
170 -18.2272720336914
182 -18.7727279663086
186 -19.0869560241699
194 -19.4347820281982
198 -19.75
};
\addlegendentry{Centralized}
\end{axis}

\end{tikzpicture}
    \vspace*{-0.2cm}
    \caption{Expected system reward $\bar{\mathcal{R}}$ versus cost $c$ for different reward functions, considering $N = 10$ sources.}
    \label{fig:10_sources_avg_reward_vs_cost}
        \vspace*{-0.2cm}
\end{figure}
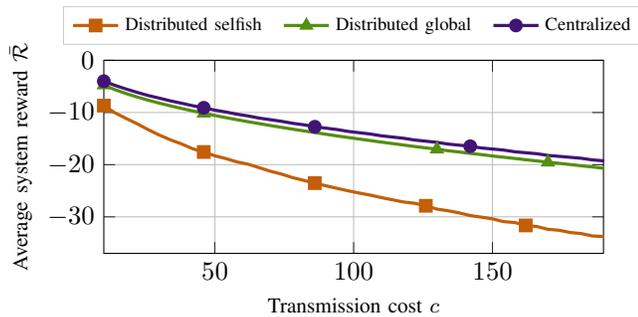

Overall, these phenomena result in the average system reward $\bar{\mathcal{R}}$ depicted in Fig.~\ref{fig:10_sources_avg_reward_vs_cost}. The \textit{centralized} policy achieves the best performance in terms of total reward, although followed quite closely by the \textit{distributed global} policy. The latter leads to similar update probabilities and costs, but pays for its inefficiencies related to a distributed management. Finally, the \textit{distributed selfish} policy ends up performing quite poorly, due to its myopic overestimation of the transmission probabilities. This is motivated by the fact that the lack of coordination eventually leads to multiple concurrent updates by the sources, which effectively represents a waste of resources.

This poor performance of the distributed selfish policy may be regarded as surprising, at least in part. Even though \cite{badia2021age} argued that the \gls{PoA} of multiple uncoordinated sources is non-negligible even in the absence of an explicit competition (here, all sources try to update the same process of interest), the reason for the inefficiency seems to be more related to the distributed management rather than to the lack of awareness of the system state. Indeed, a stateful optimization does not improve the situation and possibly may even make it worse. This is actually a known counterintuitive conclusion of many game theoretic approaches, most notably those involving Stackelberg games \cite{canzian2012promoting}, where the increase of information gained by a selfish player does not necessarily improve the system performance, as selfish players try to use this additional knowledge to their own advantage and not to the system's.
This may actually be a problem in massive \gls{iot} networks \cite{chen2020age} and will possibly prompt more theoretical investigations on how to remove this inefficiency and compensate the \gls{PoA}, so as to obtain a better \gls{aoi}-aware management of multiple sources in pervasive scenarios.
On a positive note, the \textit{distributed global} policy incurs a negligible performance degradation compared to the \textit{centralized} one, thus showing that a distributed system where the sources are incentivized to cooperate can achieve near-optimal efficiency despite the lack of coordination.

\section{Conclusions and Future Work}
\label{sec:concs}
We considered a scenario with multiple sensing nodes trying to update a single value of \gls{aoi} at the receiver's side. Even when medium access intricacies are blurred and an idealized collision-free distributed scheme is approached, the lack of coordination of terminals may be harmful for the system \gls{aoi}. This result was proven even under a stateful optimization policy, in which all nodes are aware of the value of this \gls{aoi}, and desire to minimize it. However, if they incur a cost in sending updates and act in a distributed fashion without explicit competition, but just individually following a selfish objective, the overall result is a high \gls{PoA} \cite{prospero2021resource}.

Extensions of this analysis include, from an engineering standpoint, the investigation of more specific access protocols \cite{munari,badia2022game} %,yavascan2021analysis} 
to see whether the presence of collisions (and schemes to recover from collisions) confirm, mitigate, or even amplify this problem identified for an idealized scenario where multiple transmissions never collide \cite{badia2021age}.

Further game theoretic investigations can explore how to improve the network management by establishing explicit collaborations among the players. An often exploited game theoretic narrative seeks to establish collaboration through dynamic iterations, without any explicit desire for cooperation but just in accordance to the individual objectives of the selfish players \cite{sorin1997cooperation}. It is even possible that explicit rewarding mechanisms are foreseen and implemented \cite{hu2020blockchain}.
In our opinion, these can be practical ways to achieve network efficiency in large scale systems, and as such should be certainly pursued.

\balance
\bibliographystyle{IEEEtran}
\bibliography{IEEEabrv,bibl}

\end{document}